\documentclass[aps,prl,twocolumn,amsfonts,nobibnotes,superscriptaddress,showpacs]{revtex4-1}

\usepackage{dcolumn}% Align table columns on decimal point
\usepackage{amsmath}
\usepackage{amssymb}
\usepackage{graphicx}
\usepackage{bm}
\usepackage{mathrsfs}
\usepackage{color}

\begin{document}

\title{Anomalous Anisotropic Exciton Temperature Dependence in Rutile TiO$_2$}

\vspace{2cm}

\author{Edoardo Baldini}
	\affiliation{Laboratory of Ultrafast Spectroscopy, ISIC and Lausanne Centre for Ultrafast Science (LACUS), \'Ecole Polytechnique F\'{e}d\'{e}rale de Lausanne, CH-1015 Lausanne, Switzerland}

\author{Adriel Dominguez}
	\affiliation{Max Planck Institute for the Structure and Dynamics of Matter, Hamburg, Germany}

\author{Letizia Chiodo}
	\affiliation{Unit of Nonlinear Physics and Mathematical Modeling, Department of Engineering, Universit\`a Campus Bio-Medico di Roma, Via \'Alvaro del Portillo 21, I-00128, Rome, Italy}
	
\author{Evgeniia Sheveleva}
	\affiliation{Department of Physics, University of Fribourg, Chemin du Mus\'ee 3, CH-1700 Fribourg, Switzerland}

\author{Meghdad Yazdi-Rizi}
	\affiliation{Department of Physics, University of Fribourg, Chemin du Mus\'ee 3, CH-1700 Fribourg, Switzerland}
	
\author{Christian Bernhard}
	\affiliation{Department of Physics, University of Fribourg, Chemin du Mus\'ee 3, CH-1700 Fribourg, Switzerland}
	
\author{Angel Rubio}
	\affiliation{Max Planck Institute for the Structure and Dynamics of Matter, Hamburg, Germany}
	\affiliation{Departamento F\'isica de Materiales, Universidad del Pa\'is Vasco, Av. Tolosa 72, E-20018, San Sebasti\'an, Spain}
	
\author{Majed Chergui}
	\affiliation{Laboratory of Ultrafast Spectroscopy, ISIC and Lausanne Centre for Ultrafast Science (LACUS), \'Ecole Polytechnique F\'{e}d\'{e}rale de Lausanne, CH-1015 Lausanne, Switzerland}

\date{\today}

\begin{abstract}
Elucidating the details of the electron-phonon coupling in semiconductors and insulators is a topic of pivotal interest, as it governs the transport mechanisms and is responsible for various phenomena such as spectral-weight transfers to phonon sidebands and self-trapping. Here, we investigate the influence of the electron-phonon interaction on the excitonic peaks of rutile TiO$_2$, revealing a strong anisotropic polarization dependence with increasing temperature, namely an anomalous blueshift for light polarized along the a-axis and a conventional redshift for light polarized along the c-axis. By employing many-body perturbation theory, we identify two terms in the electron-phonon interaction Hamiltonian that contribute to the anomalous blueshift of the a-axis exciton. Our approach paves the way to a complete \textit{ab initio} treatment of the electron-phonon interaction and of its influence on the optical spectra of polar materials.
\end{abstract}

%\keywords{}

\maketitle

The temperature (T) dependence of elementary excitations is a central subject in condensed matter physics, as it provides insightful information on the microscopic details of many-body interactions and correlations. To this end, over the past five decades, considerable efforts have been devoted to studying the T-effects on the optical spectra of materials, where elementary excitations in the long-wavelength regime possess a clear spectroscopic fingerprint. In this regard, an old topic is represented by the T-dependence of interband transitions and excitons in standard band semiconductors and insulators~\cite{cardona2001renormalization, giustino}. The energy of these excitations ($E\mathrm{_{exc}}$) typically undergoes a sizeable softening with increasing T, but in a few exceptional cases, the opposite effect or more complex T dependences have been observed~\cite{keffer1968pbte, yu1973anomalous, rossle2013optical}. Part of this renormalization is accounted for by the thermal expansion of the lattice, but the major contribution arises from the structure of the electron-phonon interaction (EPI). \\
\indent To model the measured dependences of $E\mathrm{_{exc}}$, simple algebraic expressions have been initially used, the most common of which is the empirical Varshni law~\cite{varshni1967temperature}. More accurate fits were obtained by using Bose-Einstein statistical factors with average acoustic and optical phonon frequencies, an approach that finds theoretical justification in pseudopotential theory \cite{allen1983temperature}. Within this framework, anomalous T-dependences of $E\mathrm{_{exc}}$ can be described by assuming that the contributions due to phonons with low- and high frequencies retain opposite signs \cite{vina1984temperature}. A more rigorous generalization of this approach, using a distribution of phonon energies, was proposed \cite{collins1990indirect}, in which $E\mathrm{_{exc}}(T)$ can be described as

\begin{equation}
E\mathrm{_{exc}}(T) = E\mathrm{_{0}} - \int d\omega f(\omega) \Big[ n\mathrm{_{BE}}(\omega,T) +\frac{1}{2} \Big] - E\mathrm{_{th}}(T)
\end{equation}
where $E\mathrm{_0}$ is the energy gap at zero T, $n\mathrm{_{BE}}$ is the Bose-Einstein statistical factor (e$^{\hbar\omega/k\mathrm{_B}T}$ - 1)$^{-1}$, f($\omega$) is a weighing factor and the last term $E\mathrm{_{th}}(T)$ accounts for the lattice thermal expansion. The weighing factor f($\omega$) can be decomposed into a product of the phonon density of states (PDOS) and a factor related to the EPI strength. However, this method suffers from intrinsic complexity, requiring detailed knowledge of the measured/calculated PDOS and EPI constants. Approximated models have been employed in isotropic materials where the PDOS is characterized by van Hove singularities associated with specific phonon modes \cite{cardona2014temperature}. In summary, the strength of these models lies on their ability to reveal the gross features of the EPI, albeit at a phenomenological level. As a result, these methods lose track of the microscopic details of the EPI, for which a full \textit{ab initio} treatment is needed. A step further in this respect involves the description of different sources contributing to the EPI in materials with an intrinsic degree of optical anisotropy.

\begin{figure*}[t]
	\begin{center}
		\includegraphics[width=2\columnwidth]{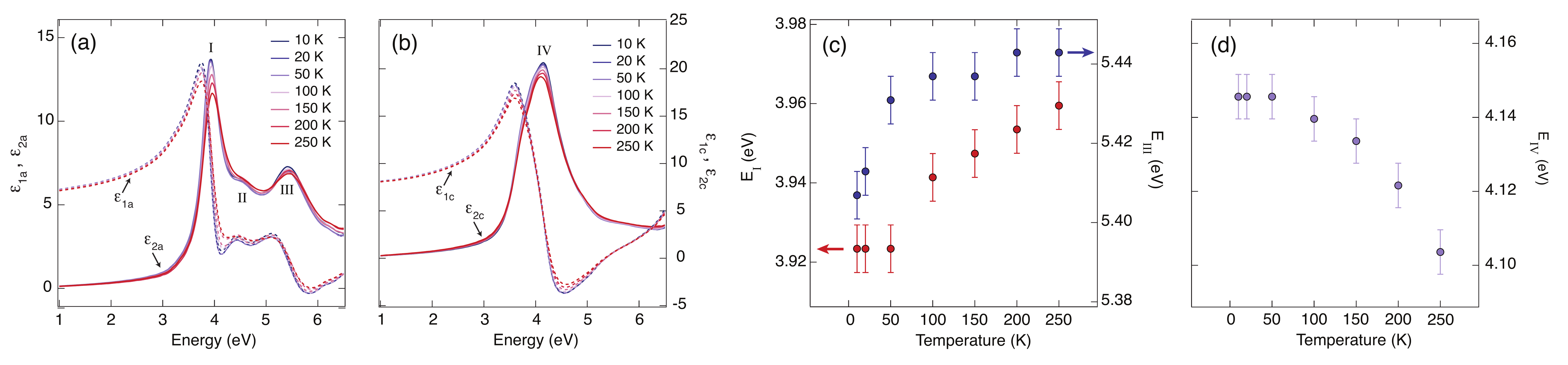}
		\caption{Measured T dependence of the dielectric function of a rutile TiO$_2$ single-crystal in (a) a-axis and (b) c-axis polarization. Behaviour of the peak energy for the charge excitations (c) I and III and (d) IV as a function of T.}
		\label{fig:Fig1}
	\end{center}
\end{figure*}

In this Letter, we perform T-dependent spectroscopic ellipsometry (SE) measurements on the polar insulator rutile TiO$_2$. To our knowledge, this is the first study to report the low-T spectra of this material and it reveals an anomalous anisotropic T-dependence of its resonant excitons. By applying state-of-the-art many-body perturbation theory calculations, we go beyond the established phenomenological models, identifying two terms of the EPI Hamiltonian which lead to the exciton hardening for increasing T. Our study paves the way to a complete quantitative treatment of the EPI in strongly interacting semiconductors and insulators, an aspect of pivotal importance for the design of future electrical and optoelectronic devices.

The SE measurements were performed on a (010)-oriented rutile TiO$_2$ single crystal. SE provides significant advantages over conventional reflection methods: (i) it is self-normalizing and does not require reference measurements; (ii) $\epsilon_1$($\omega$) and $\epsilon_2$($\omega$) are obtained directly without a Kramers-Kronig (KK) transformation; (iii) in the ultraviolet, SE is less affected by the surface roughness of the sample than normal-incidence reflectivity. Many-body perturbation theory at the level of the GW and the Bethe-Salpeter Equation (BSE) \cite{ref:hedin1, ref:onida} was employed to compute the band structure and the dielectric response of rutile TiO$_2$. Details of the experimental methods and the \textit{ab initio} calculations are reported in the Supplementary Material (SM) section.

Figures 1(a,b) show the spectra of the real and imaginary parts of the dielectric function, $\epsilon_1$($\omega$) and $\epsilon_2$($\omega$), along the a- and c-axis, respectively, at different T's. As expected, the substantial difference between the lattice constants a = 4.59 \AA~ and c = 2.96 \AA~results in a strong anisotropy of the optical properties. The low-T $\epsilon_2$($\omega$) spectra along the a-axis (Fig. 1(a)) are dominated by a narrow excitation at 3.93 eV (I), followed by a weaker shoulder at 4.51 eV (II) and a broader feature at 5.42 eV (III). This allows us to resolve the presence of feature II, which disappears for increasing T. In contrast, all the other excitations are clear cut at 250 K. The c-axis spectra (Fig. 1(b)) consist instead of a single broad feature peaking at 4.15 eV (IV). The T-evolution of peak energies I and III is shown in Fig. 1(c), while that of peak IV is in Fig. 1(d). Remarkably, we observe a large qualitative difference in the T behaviour of the excitations. Transitions I and III along the a-axis display a sizeable blueshift of 36 $\pm$ 6 meV with increasing T, while excitation IV along the c-axis undergoes an opposite redshift of 42 $\pm$ 6 meV. To our knowledge, this is the first example of a band insulator showing an opposite T behaviour of the excitations along the two polarization channels. In rutile TiO$_2$, the thermal expansion along both axis has a regular T dependence and thus should contribute to a softening of the optical transition energies as the lattice expands with increasing T~\cite{rao1970thermal}. Thus, the evolution of peak I and III with T is anomalous and is likely related to peculiar effects of the EPI at finite T. 

To rationalize our data, we present \textit{ab initio} calculations both at zero and finite T, including many-body electron-hole correlations and the effect of the EPI, on top of Density-Functional Theory results. We first compute $\epsilon_2$($\omega$) at zero T with and without many-body electron-hole correlations. Figures 2(a,b) compares the SE data at 10 K (blue lines) with the optical spectra in the uncorrelated-particle picture (red lines) obtained within the random-phase approximation (RPA) on top of GW, and the many-body optical spectra (violet lines) calculated by solving the BSE (see SM for computational details). As previously reported, only the inclusion of many-body correlations leads to the correct description of the experimental data~\cite{lawler2008optical, chiodo2010self, kang2010quasiparticle, landmann2012electronic}. However, already at this stage, the present combined experimental-theoretical effort has two clear advantages over previous studies: (i) the experimental $\epsilon_2$($\omega$) is measured directly via SE at 10 K, in contrast with the one extracted by a KK analysis at 300 K \cite{cardona1965optical}, and (ii) our GW-BSE spectra are calculated with a higher degree of convergence than previously \cite{lawler2008optical, chiodo2010self, kang2010quasiparticle, landmann2012electronic}, using a fine $k$-point grid of 16$\times$16$\times$20 and including 10 valence bands (VBs) and 10 conduction bands (CBs). As a consequence, we get an excellent agreement between the low-T SE spectra and the BSE calculations. Along the a-axis, the sharp absorption maximum at 3.99 eV lies very close to band I (3.93 eV). A shoulder emerges around 4.57 eV, which clearly corresponds to feature II (4.51 eV). This excitation was previously not resolved in either the experimental data (obscured at 300 K) \cite{cardona1965optical, tiwald2000measurement} or in the theoretical spectra (due to the lower convergence) \cite{lawler2008optical, chiodo2010self, kang2010quasiparticle, landmann2012electronic}. Finally, a transition at 5.37 eV is also apparent, corresponding to the experimental peak III (5.42 eV). Along the c-axis, a doublet structure appears, whose centre of mass at 4.24 eV can be associated with the experimental peak IV (4.15 eV). Importantly, all these excitations lie above the direct VB-to-CB optical transition evaluated at the GW level (3.34 eV, indicated by a dashed vertical line in Figs. 2(a,b)) and can therefore be described as resonant excitons. Our calculations also find a bound exciton at 3.19 eV along both axes. However, it is optically dark and arises from transitions between the VB maximum and the CB minimum at the $\Gamma$ point of the Brillouin zone. The detailed real and reciprocal space analysis of all the optical excitations is presented in \S~II.B of the SM. 

\begin{figure}[t]
\begin{center}
\includegraphics[width=0.95\columnwidth]{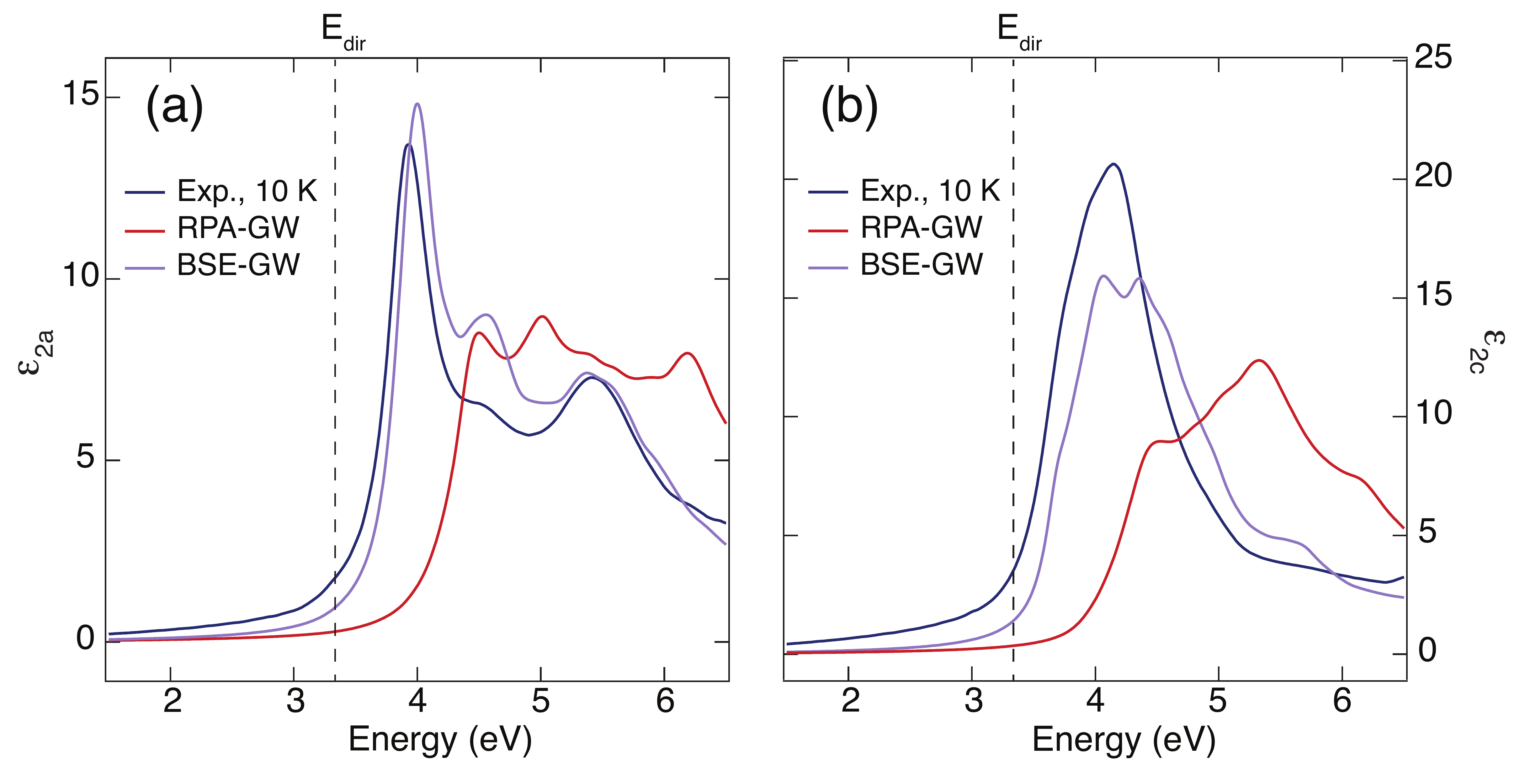}
\caption{(a-b), Calculated imaginary part of the dielectric function at 10 K with the electric field polarized along (a) the a-axis and (b) the c-axis. The experimental data are shown in blue, the calculated spectra in the RPA-GW scheme at zero T in red and the calculated spectra in the BSE-GW scheme at zero T in violet. The quasiparticle direct gap E$\mathrm{_{dir}}$ = 3.34 eV is indicated by a dashed vertical line.}
\label{fig:Fig2}
\end{center}
\end{figure}

We now address the observed anomalous T-behaviour of excitons I and III and identify its possible sources. Capturing the T-dependence of the exciton peaks requires to go beyond the zero T and frozen lattice approximations, including the zero-point renormalization (ZPR) as well as effects of finite T. This becomes a formidable task, as there is a plethora of ways the electrons interact with the lattice degrees of freedom in a crystal \cite{yucardona, giustino}. An assumption that is usually made involves the truncation of the electron-phonon perturbation theory series after the second-order terms \cite{cardona2001renormalization}. Within this approximation, the most important contribution is the effect of the first-order EPI Hamiltonian to second-order in perturbation theory (the so-called Fan-Migdal terms). For a simple semiconductor with parabolic and non-degenerate VB and CB, the Fan-Migdal matrix elements lead to the well known ``Varshni effect" \cite{varshni1967temperature}, namely a redshift of the bandgap with increasing T (note that more complex electronic band structures might very occasionally lead to a blueshift). Depending on the details of the electron-phonon matrix element, different effects arise \cite{yucardona}. In the long-wavelength limit, transverse acoustic (TA) and longitudinal acoustic (LA) phonons typically couple to the electrons via the deformation potential and the piezoelectric interactions, while transverse optical (TO) and longitudinal optical (LO) modes couple via the deformation potential interaction only. An additional contribution to the first-order EPI Hamiltonian at \textbf{q} $\sim$ 0 arises in polar or partially ionic materials, since polar LO phonons can yield a macroscopic polarization, described in terms of the Fr\"ohlich interaction \cite{frohlich1950xx}. Beyond Fan-Migdal terms, also the effect of second-order EPI in first-order perturbation theory (the so-called Debye-Waller or Yu-Brooks terms) have been demonstrated to provide a non-negligible contribution \cite{allen1981theory}.

\begin{figure}[b]
	\begin{center}
		\includegraphics[width=0.95\columnwidth]{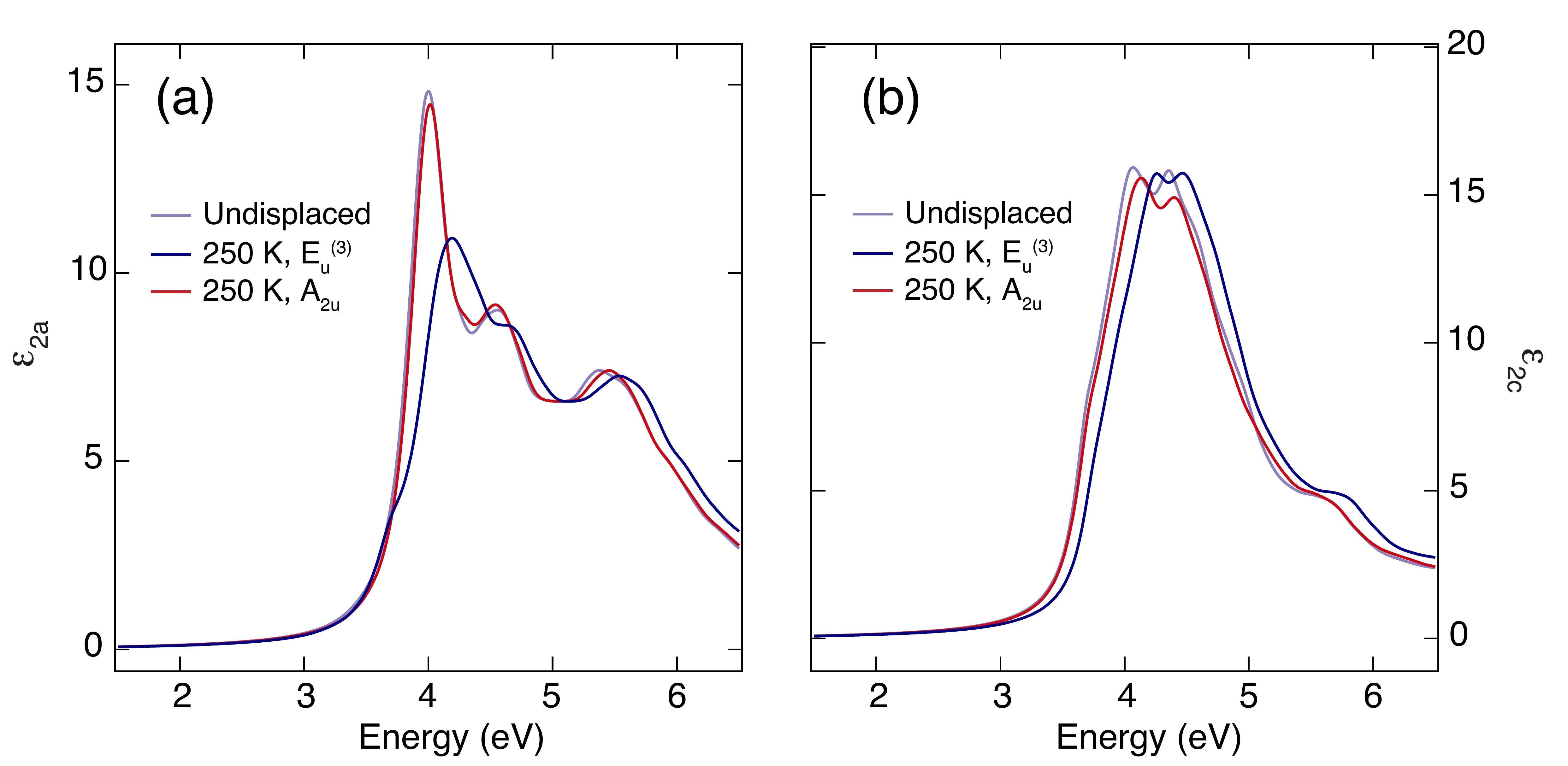}
		\caption{(a-b), Imaginary part of the dielectric function at 10 K with the electric field polarized along (a) the a-axis and (b) the c-axis. The calculated spectra in the BSE-GW scheme for the undisplaced unit cell at T = 0 are shown in violet, the calculated spectra for the unit cell displaced along the $E\mathrm{_u}^{(3)}$ phonon mode at 250 K in blue, and the calculated spectra for the unit cell displaced along the $A\mathrm{_{2u}}$ phonon mode at 250 K in red.}
		\label{fig:Fig3}
	\end{center}
\end{figure}

A complete analysis of the EPI requires to assess the impact of ZPR and T effects on the elementary charge excitations. As far as the ZPR is concerned, recent theoretical calculations on rutile TiO$_2$ estimated a decrease of 150 meV for the zero T single-particle gap and predicted its blueshift as a function of T (in contrast to the redshift shown by other insulators)  \cite{monserrat2016correlation, giustino}. Unraveling the role of T is instead complicated by the presence of a higher degree of complexity compared to conventional isotropic and non-polar materials. We first rule out any involvement of piezoelectric coupling, since rutile TiO$_2$ belongs to the $D\mathrm{_{4h}}$ space-group. On the contrary, a significant contribution is expected from the deformation of the electronic potentials due to the atomic displacements. Its sign is determined by the lattice structure and the electronic states forming the VB and the CB. Its magnitude depends on the amplitude of the atomic displacement $u$, which in the harmonic approximation is related to the atomic effective mass $\mu$, the eigenfrequency of the phonon mode $\omega$ and the occupation factor $n_\mathrm{{BE}}$, according to $<u^2>$ = $\hbar$(1 + 2$n\mathrm{_{BE}}$)/ 2$\mu\omega$ where $<...>$ means a thermal average. The resulting shift of an exciton/interband transition energy ($E_\mathrm{{exc}}$) is nearly constant at $k\mathrm{_B} T \ll \hbar\omega$ where it is dominated by the quantum lattice fluctuations, it starts deviating around $k\mathrm{_B} T \sim \hbar\omega$, and is proportional to T at $k\mathrm{_B} T \gg \hbar\omega$. To account for the contribution of the deformation potential to the EPI at finite T, we calculate the single- (GW) and two-particle (BSE) excitation spectra when the ions in the primitive unit cell are displaced statically according to specific eigenvectors of interest. Similarly to the case of anatase, also in rutile TiO$_2$ the high-frequency LO $E\mathrm{_{u}}^{(3)}$ mode at 829.6 cm$^{-1}$ and the $A\mathrm{_{2u}}$ mode at 796.5 cm$^{-1}$ are expected to possess the strongest coupling to the electronic degrees of freedom, at least in the a-axis response \cite{ref:deskins, moser2013tunable, ref:baldini_TiO2}. We estimate their deformation potentials to provide a possible explanation for the observed anomalous exciton blueshifts. We find that: i) all excitons strongly react to a unit cell displacement along the eigenvector of the $E\mathrm{_{u}}^{(3)}$ mode from zero-T to 250 K, showing a pronounced blueshift of 190 meV along the a-axis and of 130 meV along the c-axis (blue curves); ii) the a-axis excitons are barely blueshifted ($\sim$ 20 meV) by a unit cell displacement along the eigenvector of the $A\mathrm{_{2u}}$ mode, while peak IV shows a sizeable blueshift of 60 meV (red curves). As a result, we conclude that the deformation potential interaction between the $E\mathrm{_{u}}^{(3)}$ LO phonon and the in-plane charge excitations is so strong that it can account for part of the unconventional blueshift displayed by excitons I and III in the experimental spectra.

Since rutile TiO$_2$ is a polar material, also the Fr\"ohlich interaction is expected to play a major role \cite{frohlich1950xx}. To quantify its influence on the charge excitations, we recall that the energy shift induced by this interaction is \cite{frohlich1950xx, fan}
\begin{equation}
\Delta E\mathrm{_{exc,\nu}} = -A_{\nu}(\epsilon_{\infty,\nu}^{-1} - \epsilon_{0,\nu}^{-1})(1+2n\mathrm{_{BE}})
\end{equation}
\noindent where $\nu$ = ($a$,$c$) depending on the crystal axis, $\epsilon_{\infty,\nu}$ and $\epsilon_{0, \nu}$ are the dielectric constants at energies well above and below the phonon range, respectively. $A_{\nu}$ is a nearly T-independent prefactor that reads
\begin{equation}
A_{\nu} = e^2 \sum_i \sqrt{\hbar \omega\mathrm{_{LO, i}}} \epsilon_{\infty,\nu} \Bigg(\frac{\sqrt{2m_e^*}}{\hbar} + \frac{\sqrt{2m_h^*}}{\hbar}\Bigg)
\end{equation}
where $e$ is the fundamental charge and $m\mathrm{_e^*}$ ($m\mathrm{_h^*}$) is the electron (hole) effective mass. The sum runs over all the $i$ polar LO phonons with frequency $\omega_\mathrm{{LO}}$, which in rutile TiO$_2$ are represented by the $E\mathrm{_u^{(1)}}$, $E\mathrm{_u^{(2)}}$, $E\mathrm{_u^{(3)}}$ and $A\mathrm{_{2u}}$ modes \cite{traylor1971lattice}. In Eqs. (2)-(3), $\epsilon_\mathrm{{\infty,\nu}}$ and $\omega\mathrm{_{LO}}$ are nearly T-independent while $\epsilon\mathrm{_{0,\nu}}$ strongly varies with T. As a result, for $k\mathrm{_B} T \ll \hbar\omega$, where $n\mathrm{_{BE}}$ is nearly constant, the T dependence of $\Delta E\mathrm{_{exc,\nu}}$ due to the Fr\"ohlich interaction is determined by the T behaviour of $\epsilon\mathrm{_{0, \nu}}$. In rutile TiO$_2$, capacitance measurements \cite{samara1973pressure} determined $\epsilon_{0, a}$ to decrease from 115 to 90 when T is raised from 4 K to 300 K. Analogously, $\epsilon\mathrm{_{0, c}}$ reduces its value from 251 to 167 for the same T increase. From our data in Figs. 1(a,b), we establish $\epsilon\mathrm{_{\infty,a}}$ $\sim$ 5.8 and $\epsilon_{\infty,c}$ $\sim$ 8. Finally, from our \textit{ab initio} calculations, we extract the values of $m\mathrm{_e^*}$ and $m\mathrm{_h^*}$ by performing a parabolic fit of the bands involved in the transitions contributing to excitons I and IV (see SM). Substituting all values in Eq. (2) yields a blueshift of 86 meV for exciton I and of 160 meV for exciton IV.

In summary, our calculations identify that both the deformation potentials of the LO $E\mathrm{_{u}}^{(3)}$ and $A\mathrm{_{2u}}$ normal modes and the Fr\"ohlich interaction lead to a pronounced blueshift of all excitons in rutile TiO$_2$. As such, these two effects lie at the origin of the experimental shift retrieved along the a-axis, albeit the latter is smaller than predicted by our calculations. This discrepancy can be related to the simultaneous contribution of other modes producing a sizeable redshift, as well as to the action of the Debye-Waller terms of the EPI Hamiltonian, which are all neglected in our treatment. Moreover, we remark that Eq. (2) does not account for the presence of strong electron-hole Coulomb interaction. The latter is also expected to vary with T, since the ionic contribution to its screening (embodied by $\epsilon\mathrm{_{0, \nu}}$) is highly T dependent. In contrast, the blueshift predicted along the c-axis is not experimentally observed, and exciton IV undergoes a conventional redshift for increasing T. This implies that the Fan-Migdal terms of the EPI Hamiltonian contributing to the redshift of this exciton are much more efficient along the c-axis. Such a behaviour can depend on the anisotropic structure of the PDOS or on the anisotropic strength of the EPI. To explain this softening, a close inspection of the partial PDOS in rutile TiO$_2$ is needed \cite{traylor1971lattice, sikora2005ab}. While the a-axis PDOS retains a complex structure with different modes extending between 98 and 838 cm$^{-1}$, the c-axis PDOS mainly shows a lower peak at 98 cm$^{-1}$ due to TA modes and a very prominent peak at 467 cm$^{-1}$. The latter is a van Hove singularity caused by the branches of the Raman-active $E\mathrm{_g}$ mode and the polar $E\mathrm{_u^{(2)}}$ LO mode, and it is absent in the a-axis PDOS. As such, we expect the anisotropic deformation potential coupling to these modes to be the main cause behind the softening of exciton IV. This scenario can be confirmed phenomenologically by using an approximated model based on Eq. (1), which yields an excellent and robust fit only when two Bose-Einstein oscillators at 98 cm$^{-1}$ and 467 cm$^{-1}$ are imposed (see \S IV and Fig. S2 of the SM). From the fit, we obtain that the high-frequency modes have a $\sim$ 5.5 larger coupling than the one of the low-frequency mode. This indicates that the effect of the LO phonons on the T dependence of exciton IV is more important than that of the TA phonons, in accordance with the relative ratio of the peak heights ($A\mathrm{_{LO}}$/$A\mathrm{_{TO}}$ $\sim$ 5.8) in the c-axis PDOS \cite{sikora2005ab}. A complete \textit{ab initio} electron-phonon calculation, \textit{i.e.} including ZPR, T-effects and electron-electron correlations, and following some of the approaches recently introduced in Refs. \cite{monserrat2016correlation, zacharias1, zacharias2}, should confirm this trend.

In conclusion, in this work we unraveled the anisotropic evolution of the exciton peaks of rutile TiO$_2$ with T and reproduced the optical response of the material via many-body perturbation theory. From first principle calculations, we evaluated different contributions of the electron-phonon coupling that lead to an anomalous blueshift/redshift of the excitonic peaks with increasing T. Our approach paves the way to a complete microscopic treatment of the electron-phonon coupling and of its influence on the optical spectra of polar semiconductors, which is of pivotal importance for optimizing their applications.

\begin{acknowledgments}
We thank Ingalena Bucher for her contribution during the measurements. We acknowledge financial support from the Swiss NSF via the NCCR:MUST and the contracts No. 206021 157773, 20020 153660 and 407040 154056 (PNR 70), the European Research Council Advanced Grants H2020 ERCEA 695197 DYNAMOX and QSpec-NewMat (ERC-2015-AdG-694097), Spanish Grant FIS2013- 46159-C3-1-P, Grupos Consolidados del Gobierno Vasco (IT578-13), COST Actions CM1204 (XLIC), MP1306 (EUSpec) and European Unions H2020 program under GA no.676580 (NOMAD).
\end{acknowledgments}

\newpage

\clearpage
\newpage

\setcounter{section}{0}
\setcounter{figure}{0}
\renewcommand{\thesection}{S\arabic{section}}  
\renewcommand{\thetable}{S\arabic{table}}  
\renewcommand{\thefigure}{S\arabic{figure}} 
\renewcommand\Im{\operatorname{\mathfrak{Im}}}

\section{S1. Spectroscopic ellipsometry}

We used spectroscopic ellipsometry (SE) to measure the complex dielectric function of the sample, covering the spectral range from 1.50 eV to 6.50 eV. The measurements were performed using a Woollam VASE ellipsometer. Rutile TiO$_2$ single crystals were purchased from MTI Corporation, characterized by x-ray Laue diffraction and polished down to optical grade on their (010) surface. Subsequently, the samples were mounted in a helium flow cryostat, allowing measurements from room temperature (T) down to 10 K. The T-dependent measurements were performed at $<$10$^{-8}$ mbar to prevent measurable ice-condensation onto the sample. Anisotropy corrections were performed using standard numerical procedures \cite{ref:aspnes}.

\section{S2. \textit{Ab initio} calculations}
\label{AbInitio_Details}

%Rutile TiO$_2$ crystallizes in a tetragonal unit cell, with a = b = 4.59 \AA~and c = 2.96 \AA.
Many-body perturbation theory at the level of the GW and the Bethe-Salpeter Equation (BSE) \cite{ref:hedin1, ref:hedin2, ref:onida}, as implemented within the BerkeleyGW package \cite{ref:deslippe}, was employed to compute the band structure and the dielectric response of pristine rutile TiO$_2$, on top of eigenvalues and eigenfunctions obtained from Density-Functional Theory (DFT). We used the planewave pseudopotential implementation of DFT as provided by the package Quantum Espresso \cite{giannozzi2009quantum}. The DFT calculations were performed using the generalized gradient approximation (GGA) as in the Perdew-Burke-Ernzerhof (PBE) scheme for the exchange-correlation functional. For these calculations, we used normconserving pseudopotentials \cite{ref:RRKJ} including semicore states $3s$ and $3p$ (as in Ref. \cite{ref:baldini_TiO2} for anatase TiO$_2$). We used a cutoff of 200 Ry to achieve convergence of the electronic and optical properties at the many body level. A coarse grid of 5 $\times$ 5 $\times$ 7 $k$-points was used for PBE and subsequent GW calculations.
A more dense, randomly shifted, grid of 16$\times$16 $\times$20 (with interpolation of GW energies on the 5$\times$5$\times$7 grid) was implemented for solving the BSE using Haydock method, to obtain the absorption spectra. The diagonalization of excitonic Hamiltonian was done on a randomly shifted fine grid of 10$\times$10$\times$14 $k$-points. Such diagonalization provides the analysis of Kohn-Sham orbital contributions and $k$-points contributions to excitonic wavefunctions. The excitonic wavefunctions in three dimensions were also obtained via direct diagonalization. We used 2498 conduction bands (CBs) and an energy cutoff of 46 Ry for polarisability and inverse dielectric matrix, 2498 CBs for self-energy evaluation, a cutoff energy of 50 Ry and 200 Ry for screened and bare components of self-energy operator, respectively. The 10 topmost valence bands (VBs) and the 10 lowest CBs were included in the BSE solution via Haydock, while the 7 topmost VBs and the 6 lowest CBs were used for directly diagonalizing BSE Hamiltonian for the excitons analysis.

Our study goes beyond previous experimental/computational works on rutile TiO$_2$ \cite{chiodo2010self, kang2010quasiparticle, landmann2012electronic, lawler2008optical} in that: i) a higher computational convergence is achieved; ii) a precise identification and characterization of the peaks in the optical absorption spectrum is made; iii) the T = 0 calculations can be compared with the low-T dielectric function of the material, measured directly via SE and not extracted through a Kramers-Kronig analysis; iv) T effects are included and discussed.

As far as the T = 0 calculations are concerned, the GW indirect bandgap ($\Gamma$-R) and direct bandgap ($\Gamma$-$\Gamma$) are 3.30 eV and 3.34 eV, respectively. These values are converged up to 5 meV and are in good agreement with photoemission and inverse-photoemission results \cite{tezuka1994photoemission}. The present highly-converged value is lower than the gap given in Ref. \cite{chiodo2010self}, where a smaller number of bands and $k$-points was used. The excitonic energies were subsequently obtained by diagonalizing the excitonic Hamiltonian. The first eigenvalue for both light polarizations correspond to a dark triplet exciton with energy 3.19 eV. This dark exciton has a major contribution from the transitions from the VB to the CB at the $\Gamma$ point. Its binding energy is estimated $\sim$ 150 meV. The large excitonic peak, for light polarized along the a-axis, at 3.99 eV is composed of various eigenvalues with reciprocal space contributions from a broad region around $\Gamma$. This is a resonant, non-bound exciton. For light polarized along the c-axis, the intense and broad excitonic peak centered at 4.24~eV is composed of various eigenvalues, with reciprocal space contributions from a wider region around the $\Gamma$ point and strong contributions from points close to $\Gamma$ along the $\Gamma$-Z line. Also this peak is a resonant non-bound exciton, retaining a bulk delocalized character. 
The excitonic wavefunctions on the (001) plane are represented in Fig. S1(a,b). The isosurface represents the electronic part of the excitonic squared modulus wavefunction, with the hole fixed at a site close to one oxygen atom. 

\begin{figure}[t]
	\begin{center}
		\includegraphics[width=0.95\columnwidth]{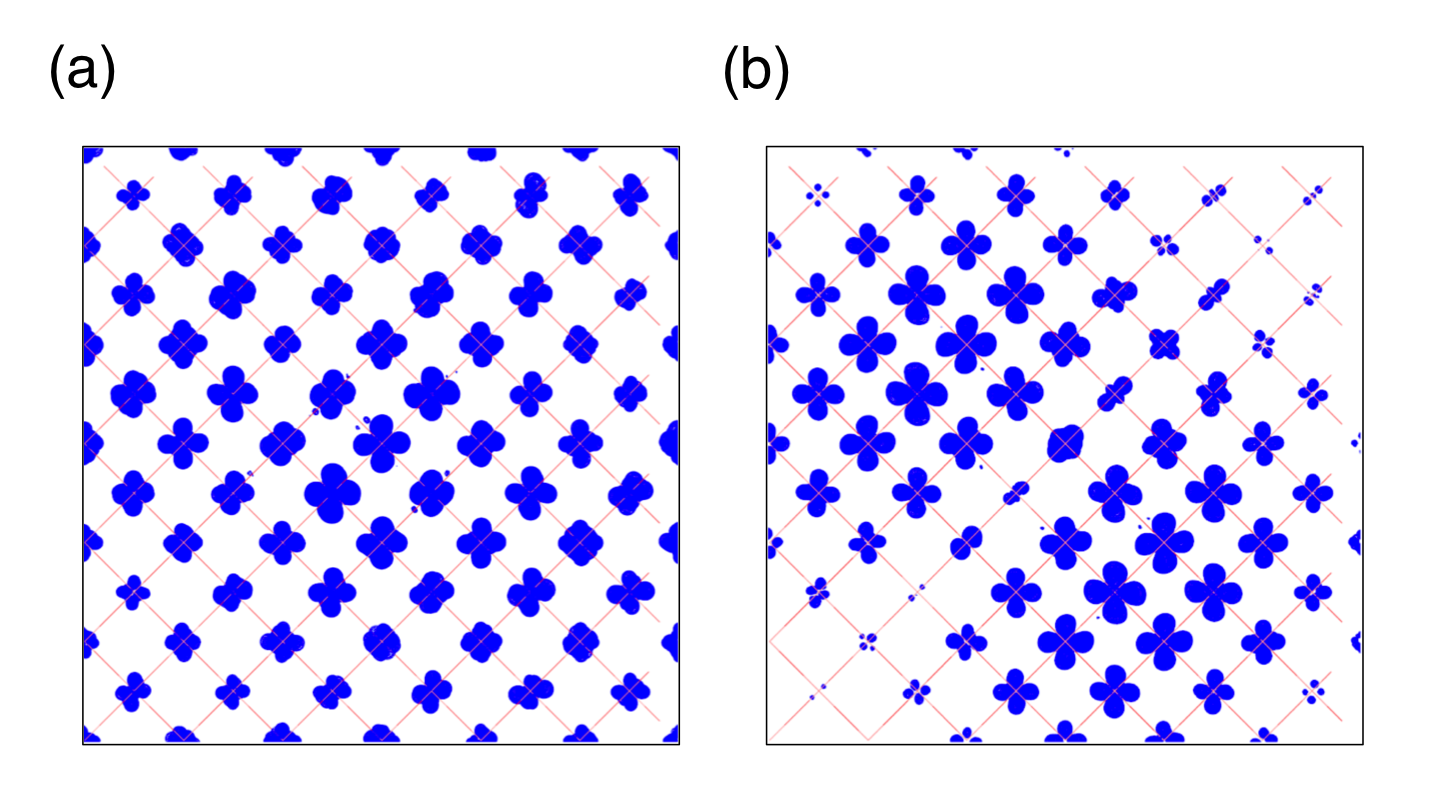}
		\caption{Wavefunctions of the fundamental charge excitations in rutile TiO$_2$. Isosurface representation of the electronic configuration on the (001) plane when the hole of the considered excitonic pair is localized close to one oxygen atom. The coloured region represents the excitonic squared modulus wavefunction. (a) Resonant exciton I at 3.99 eV. (b) Resonant exciton IV at 4.24 eV.}
		\label{fig:ExcitonWave}
	\end{center}
\end{figure}

To include the T-effects on the electronic and optical properties of rutile TiO$_2$, we estimated the role of the deformation potential coupling by performing frozen phonon GW-BSE calculations. We separately displaced the ions in the primitive unit cell according to the eigenvector of the longitudinal optical (LO) $E\mathrm{_{u}^{(3)}}$ and $A\mathrm{_{2u}}$ modes \cite{traylor1971lattice}. These phonons are expected to possess the strongest coupling with the electronic degrees of freedom along the a-axis of the crystal \cite{ref:deskins, moser2013tunable}. The displacement of atom $j$ was calculated from the harmonic oscillator mean square displacement at 250 K according to
\begin{equation}
< |u\mathrm{_{j}}(t)|^2> =\frac{\hbar(1 + 2n_\mathrm{{BE}})}{2 m\mathrm{_j} \omega}
\end{equation}
where $n_\mathrm{{BE}} = (e^{\hbar\omega/k\mathrm{_B} T} - 1)^{-1}$ is the Bose-Einstein statistical occupation factor, $k\mathrm{_B}$ is the Boltzmann constant, $m\mathrm{_j}$ is the atomic mass and $\omega$ is the phonon frequency. We corrected the GW gap values for the lattice expansion effect by using the thermal expansion coefficient reported in Ref. \cite{rao1970thermal}. In particular, the $a$ and $c$ lattice parameters of rutile TiO$_2$ increase by 0.3 $\%$ from 0 to 250 K. The inclusion of both the phonon-induced and the thermal expansion-induced effects leads to a net blueshift of the bandgap of $\sim$ 120 meV (in the case of the $E\mathrm{_{u}^{(3)}}$ mode) and 60 meV (in the case of the $A\mathrm{_{2u}}$ mode). A similar trend was recently reported in the case of the electronic gap (evaluated within the thermal lines method for electron-phonon coupling), which shows a non-monotonic behaviour with T \cite{monserrat2016correlation}. When solving the BSE on top of the T-corrected GW results, we find that, in the case of the $E\mathrm{_{u}^{(3)}}$ mode at 250 K, exciton I blueshifts by 190 meV and exciton IV blueshifts by 130 meV. In the case of the $A\mathrm{_{2u}}$ mode at 250 K, exciton I blueshifts by 20 meV and exciton IV blueshifts by 60 meV.

\section{S3. Fr\"ohlich Interaction}

In polar crystals, a major contribution to the electron-phonon interaction comes from the polarization of the lattice produced by the polar LO modes. The polarization in a unit cell, \textbf{p}, can be written as
\begin{equation}
\textbf{p} = e^* (\textbf{R$_1$} - \textbf{R$_2$})
\end{equation}
\noindent where \textbf{R$_1$} and \textbf{R$_2$} are the displacements of the positive and negative ions, respectively, and $e^*$ is the effective charge of the ions, which differs from the fundamental charge $e$ by the effect of the polarization of the ions as a result of their displacements. The structure of the polar electron-phonon interaction has been described by the theory of Fr\"ohlich \cite{frohlich1950xx}. The momentum-dependent matrix element of the interaction potential reads
\begin{equation}
\mathscr{M}(q) = \frac{1}{\sqrt{MN}}\frac{2\pi e e^*}{a^3 q} \sqrt{\frac{\hbar}{2\omega_{LO,q}}}\sqrt{n_{BE} + 1}
\end{equation}
where $M = M_1M_2/(M_1+M_2)$ is the reduced mass, $N$ is the total number of unit cells in the lattice, $a$ is the lattice constant (assuming a cubic crystal), $\omega\mathrm{_{LO,q}}$ is the frequency of the LO modes at a given momentum $q$ and $n\mathrm{_{BE}}$ is the statistical Bose-Einstein occupation factor.

The frequency of a LO mode is approximately constant as a function of $q$ and it is related to the transverse optical (TO) mode frequency $\omega\mathrm{_{TO}}$ via the Lyddane-Sachs-Teller relationship \cite{lyddane1941polar}
\begin{equation}
\mathrm{\omega_{LO}^2} = {\mathrm{\omega^2_{TO}}}(\epsilon_{0}/\epsilon_{\infty})
\end{equation}
where $\epsilon_{0}$ and $\epsilon_{\infty}$ are the dielectric constants at energies well below and above the phonon range, respectively. The frequency $\omega\mathrm{_{TO}}$ itself is related to $\epsilon_{0}$ and $\epsilon_{\infty}$ by
\begin{equation}
\omega\mathrm{^2_{TO}} = 2\pi e^{*2} \epsilon_{\infty}/Ma^3 (\epsilon_{0} - \epsilon_{\infty})
\end{equation}
The energy shift of a charge excitation induced by the Fr\"ohlich interaction is
\begin{equation}
\Delta E\mathrm{_{exc}} = -A \Bigg(\frac{1}{\epsilon_{\infty}} - \frac{1}{\epsilon_{0}}\Bigg)(1 + 2n\mathrm{_{BE}})
\end{equation}\\
\noindent where $A$ is a nearly T-independent prefactor that reads
\begin{equation}
A = e^2 \sqrt{\hbar \omega\mathrm{_{LO}}} \epsilon_{\infty} \Bigg(\frac{\sqrt{2m_e^*}}{\hbar} + \frac{\sqrt{2m_h^*}}{\hbar}\Bigg)
\end{equation}
where $m\mathrm{_e^*}$ ($m\mathrm{_h^*}$) is the electron (hole) effective mass. In this work, we evaluated this formula along the a- and c-axis of rutile TiO$_2$, accounting for all polar LO modes in the crystal and relying on the experimental values of $\epsilon_{0}$ and $\epsilon_{\infty}$. The values of $m\mathrm{_e^*}$ and $m\mathrm{_h^*}$ were estimated from our \textit{ab initio} calculations, by performing a parabolic fit at the $\Gamma$ point of the bands involved in the transitions I and IV (see Section II. B). Along the $\Gamma$-X line of the Brillouin zone, we obtain $m\mathrm{_e^*}$ = 0.61 $m\mathrm{_e}$ and $m\mathrm{_h^*}$ = 1.33 $m\mathrm{_e}$. Along the $\Gamma$-Z line of the Brillouin zone, we obtain $m\mathrm{_e^*}$ = 0.73 $m\mathrm{_e}$ and $m\mathrm{_h^*}$ = 5.20 $m\mathrm{_e}$. Substituting all values in Eq. (6) yields a blueshift of 86 meV for exciton I and of 160 meV for exciton IV. Therefore, the Fr\"ohlich interaction affects the fundamental charge excitations of rutile TiO$_2$, causing their hardening for increasing T. Indeed, the T dependence of the charge excitations is encoded in the T variation of $\epsilon_{0}$.

\section{S4. Softening of Exciton IV}

While the T dependence experienced by all in-plane charge excitations of rutile TiO$_2$ is anomalous compared to standard semiconductors and insulators, the behaviour of exciton IV is instead conventional. Indeed, in the experimental data, the energy of charge excitation IV undergoes a sizeable softening for increasing T, similarly to the dependence shown by silicon, germanium and other band semiconductors. From our theoretical calculations (see Sections II and III), we highlighted the effect produced by the DP coupling and the Fr\"ohlich interaction on exciton IV, observing a non-negligible hardening of the peak for increasing T. In rutile TiO$_2$, the effect of thermal expansion is not sufficient to explain the origin of the peak softening, since it is not strong enough to counterbalance the DP and Fr\"ohlich interactions. This implies that other terms in the electron-phonon interaction Hamiltonian are at the origin of the exciton IV softening. 

In the following, we describe the conventional redshift experienced by exciton IV by fitting the experimental T dependence with an approximated model. Although this represents a phenomenological approach, this is the most advanced model employed in literature other than \textit{ab initio} treatments. The most general approach uses a distribution of phonon energies to describe the impact of the electron-phonon interaction on the energy ($E\mathrm{_{exc}}$) of a charge excitation (i.e. interband transition or excitonic transition) \cite{collins1990indirect}. The shift of $E\mathrm{_{exc}}(T)$ can be described by
\begin{equation}
E\mathrm{_{exc}}(T) = E_{0} - \int d\omega f(\omega) \Big[ n\mathrm{_{BE}}(\omega,T) +\frac{1}{2} \Big] - E\mathrm{_{th}}(T)
\end{equation}
where $E\mathrm{_0}$ is the bare (unrenormalized) excitation energy at zero T, $n\mathrm{_{BE}}$ is the Bose-Einstein statistical factor (e$^{\hbar\omega/k\mathrm{_B}T}$ - 1)$^{-1}$, f($\omega$) is a weighting factor and the last term $E\mathrm{_{th}}(T)$ accounts for the lattice thermal expansion. The weighting factor f($\omega$) can be decomposed into a product of the phonon density of states (PDOS) $\rho$($\omega$) and a factor related to the EPI strength. For an isotropic crystal, the T dependent lattice expansion is given by
\begin{equation}
E\mathrm{_{th}}(T) =  -3B\Bigg(\frac{\partial E\mathrm{_{exc}}}{\partial p} \Bigg)_T \int_0^T \alpha(T') dT'
\end{equation}
with  $\alpha$(T) = L$^{-1}$($\partial$L/$\partial$T)$_p$, being the linear thermal expansion coefficient, B the bulk modulus and $(\partial E_g$/$\partial p$)$_T$ is the dependence of $E\mathrm{_{exc}}$ on hydrostatic pressure.

\begin{figure}[t]
	\begin{center}
		\includegraphics[width=0.65\columnwidth]{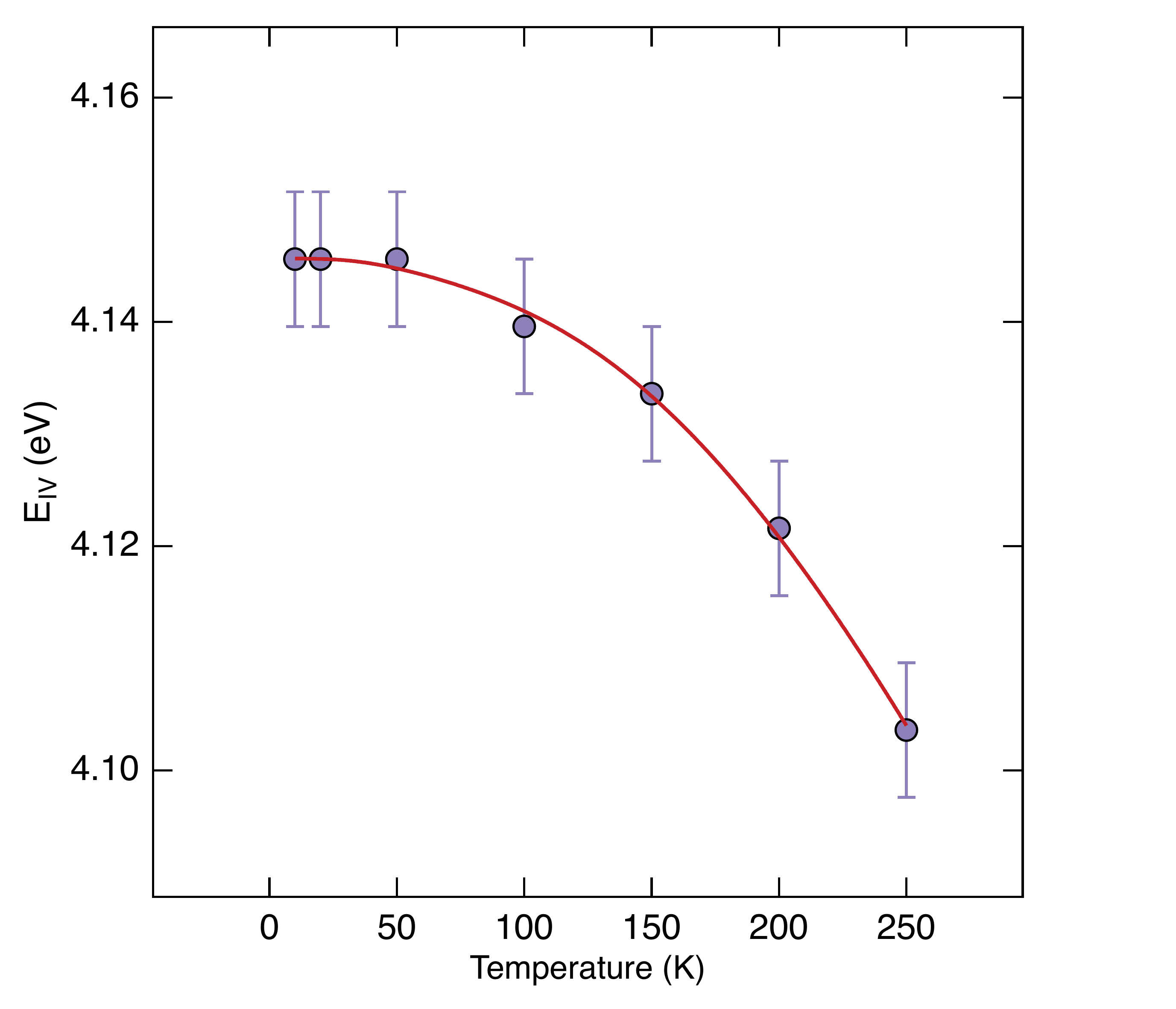}
		\caption{Temperature behaviour of the peak energy for exciton IV: experimental data are represented as violet dots, the fit is shown as a red curve.}
		\label{fig:Softening}
	\end{center}
\end{figure}

Although representing the most complete approach, Eq. (8) necessitates detailed knowledge of the PDOS and the coupling constants. Moreover, it requires significant computational efforts, making this general treatment rather expensive. This approach can be simplified when the PDOS is dominated by specific peaks due to van Hove singularities in the dispersion relation \cite{cardona2014temperature}. For example, along the c-axis, the PDOS of rutile TiO$_2$ is characterized by the presence of a peak at 12 meV (98 cm$^{-1}$) due to transverse acoustic (TA) phonons and a peak at 58 meV (467 cm$^{-1}$) due to the branches of the $E\mathrm{_g}$ and $E\mathrm{_u^{(2)}}$ phonons \cite{sikora2005ab}. As such, the PDOS $\rho$($\omega$) can be approximated by two delta-functions positioned at average TA and LO frequencies
\begin{equation}
\rho(\omega) = \delta(\omega-\omega_\mathrm{{TA}}) + \delta(\omega-\omega_\mathrm{{LO}})
\end{equation}
We assume that the coupling constant for TA phonons is  proportional to $q^2$ whereas for LO phonons is independent of $q$. Then, the T dependence of $E\mathrm{_{exc}}(T)$ can be written as
\begin{equation}
E\mathrm{_{exc}}(T) \approx E_0 - \sum_{i = 1}^2 A_i \omega_\mathrm{i} (2n_{BE} + 1) + E\mathrm{_{th}}(T)
\end{equation}
where $\omega_1$ = $\omega_\mathrm{{TA}}$ and $\omega_2$ = $\omega_\mathrm{{LO}}$. In the fit, the coupling constants A$_\mathrm{i}$ are allowed to change sign and magnitude. Conversely, the phonon frequencies $\omega_\mathrm{i}$ are maintained fixed. Moreover, the lattice expansion term $E\mathrm{_{th}}(T)$ is generalized to the case of an anisotropic (tetragonal) crystal, using the expressions for the linear thermal expansion coefficients of rutile TiO$_2$ as in Ref. \cite{rao1970thermal}. The value of the bulk modulus is taken from Ref. \cite{manghnani1969elastic}, while the dependence of on hydrostatic pressure is extracted from Ref. \cite{yin2010effective}. The results of the fit are shown in Fig. S2 as a red curve. We observe that the fit reproduces the experimental T dependence in an excellent way. Importantly, we also underline that our fit is robust with respect to the number of Bose-Einstein oscillators used and the choice of the phonon energies at 12 meV and 58 meV. Indeed, the use of different phonon energies other than 12 meV and 58 meV worsen the fit significantly. This proves that the modes playing the major role in the T renormalization of exciton IV are those displaying the strongest peaks in the c-axis PDOS. Furthermore, from the fit, we obtain two very interesting and consistent results: i) the bare excitation energy at zero T ($E\mathrm{_0}$) is found at 4.32 eV, which is 170 meV larger than the experimental excitation energy at 10 K. As such, this value is in very good agreement with the zero-point renormalization of 150 meV calculated in a recent \textit{ab initio} study of the single-particle excitation spectrum of rutile TiO$_2$ \cite{monserrat2016correlation}; ii) the modes at 58 meV have a coupling that is by about a factor of 5.5 larger than the one of the 12 meV phonons. This indicates that the effect of the high-energy LO phonons on the T dependence of exciton IV is more important than that of the low-energy TA phonons, in accordance with the relative ratio of the peak heights ($A\mathrm{_{LO}}$/$A\mathrm{_{TO}}$ $\sim$ 5.8) in the calculated c-axis PDOS. Despite the excellent parameters retrieved from our fit, we remark that this phenomenological approach neglects the detailed $q$-dependence of the electron-phonon coupling strength. However, prior to our computational study, this approximate model represented one of the most advanced treatments presented in literature.

\bibliography{Papers}
\end{document}